\documentclass[10pt,aps,prl,twocolumn,preprintnumbers,amsmath,amssymb,floatfix]{revtex4-1}
\usepackage[latin1]{inputenc}
\usepackage[T1]{fontenc}
\usepackage[english]{babel}
\usepackage[pdftex]{graphicx}
\usepackage{amsmath}
\usepackage{times}
\usepackage{psfrag}
\usepackage{csquotes}
\usepackage[usenames,dvipsnames,svgnames,table]{xcolor}
\usepackage[colorlinks,plainpages=false,linkcolor=blue,urlcolor=blue,citecolor=blue,pdfpagemode=UseNone]{hyperref}

\renewcommand{\AA}{\text{\r{A}}}

\begin{document}

\title
{
Atomic-scale detection of magnetic impurity interactions in bulk semiconductors
}

\author{Benjamin Geisler}
\email{benjamin.geisler@uni-due.de}
\author{Peter Kratzer}

\affiliation{
Fakult\"at f\"ur Physik and Center for Nanointegration (CENIDE), Universit\"at Duisburg-Essen, 47048 Duisburg, Germany
}

\date{\today}

\begin{abstract}
We demonstrate on the basis of \textit{ab initio} simulations how passivated semiconductor surfaces can be exploited to study \textit{bulklike} interaction properties and wave functions of magnetic impurities on the atomic scale with conventional and spin-polarized scanning tunneling microscopy.
By applying our approach to the case of $3d$ transition metal impurities close to the H/Si$(111)$ surface, we show exemplarily that their wave functions in Si are less extended than for Mn in GaAs, thus obstructing ferromagnetism in Si.
Finally, we discuss possible applications of this method to other dilute magnetic semiconductors.
\end{abstract}

\maketitle

Identification of materials systems to be used as dilute magnetic semiconductors (DMSs) 
has so far been following a trial-and-error approach, resulting in slow progress of this field, which is still dominated by Mn-doped GaAs~\cite{Ohno-MnGaAs:96}. 
For rational materials design, understanding the interactions in a DMS on a fundamental, atomic level is a prerequisite. 
This comprises both impurity-host and impurity-impurity interactions that are responsible for the formation of local magnetic moments and the emergence of collective order between these moments, respectively.
In this Rapid Communication we promote (preferentially spin-polarized) scanning tunneling microscopy (SP~STM) as a powerful method
to extract atomic-scale information about the magnetic properties of impurities in semiconductors. 
The results of large-scale \textit{ab initio} simulations put us in the position to devise a generally applicable experimental strategy
that exploits passivation of the dangling bonds at semiconductor surfaces
(preferentially cleavage planes)
in order to preserve the \textit{bulklike} behavior of subsurface impurities. 
Computer simulations for transition metals (TMs), such as Cr, Mn, and Fe, in the most common semiconductor, Si, serve as a proof of principle.
In advantage over conventional spectroscopic techniques, the proposed strategy using SP~STM provides a visual explanation why ferromagnetism is more difficult to achieve in Si than in GaAs. 
Moreover, our calculations show that the resolution of SP~STM is sufficiently high to distinguish different magnetic states of impurity pairs, and even to quantify their exchange interaction with the help of an externally applied magnetic field.

Most experimental methods sufficiently sensitive to detect the magnetic moments of impurities, such as electron para\-magnetic resonance or x-ray magnetic circular dichroism~\cite{Ney:08, NeyPRB:10}, average over a finite sample volume and thus lack spatial resolution. 
For instance, TM atom clusters could mask the signal from isolated impurities.  
In contrast, local information can be deduced precisely on the atomic scale via state-of-the-art (cross-sectional) STM; see, e.g., the ample literature on subsurface Mn impurities in GaAs~\cite{KoenraadFlatte:11, Yakunin:07, Kitchen:06, Garleff:08, Garleff:10}.
SP~STM is used nowadays to explore the magnetic properties of TM atoms at (semiconductor) surfaces~\cite{Wiesendanger:10, Wiesendanger:12, Loth:10, Wiesendanger-RevModPhys:09}.
However, these experiments actually map the \textit{surface} behavior of impurities, which may differ strongly from their behavior in bulk material due to hybridization of impurity states and host surface states~\cite{Jancu:08} and charge transfer 
involving unsaturated host surface bonds.

We performed spin-polarized density functional theory~\cite{KoSh65} (DFT) calculations
for $3d$ TM impurities in different structurally optimized Si systems
within the ultrasoft pseudopotential~\cite{Vanderbilt:1990,PWSCF}
and the projector augmented wave~\cite{PAW:94,USPP-PAW:99} frameworks,
together with the semilocal Perdew-Burke-Ernzerhof functional (PBE)~\cite{PeBu96}
and the hybrid functional of Heyd, Scuseria, and Ernzerhof (HSE06)~\cite{HSE:06}, respectively.
We obtained bulk results from cubic $216$-atom Si supercells ($5\times5\times5$ $\vec{k}$-point grid including $\Gamma$ for Brillouin zone sampling).
For the STM simulations, very large supercells had to be used due to the high sensitivity of wave functions to the boundary conditions
[Si$(111)$-$(9 \times 9)$ slab with $972$
atoms, two passivating H layers, $20\ \AA$~vacuum region, $2\times2\times1$ $\vec{k}$-point grid including~$\Gamma$].  
Constant current STM images were simulated subsequently in the spirit of Tersoff and Hamann~\cite{tersoff:83}
as isosurfaces of the integrated local density of states:
$\varrho(\vec{r}) = \int_{0}^{eV} \vert \text{d} \varepsilon \vert \, \sum_{n \vec{k}} \vert \psi_{n \vec{k}} (\vec{r}) \vert^2 \ \delta(\varepsilon - \varepsilon_{n \vec{k}} + E_{\text{F}})$,
$\left\lbrace z(x,y) = z \, : \, \varrho(x,y,z) = \varrho_c \right\rbrace$.
SP~STM images are differences $z_{\uparrow} - z_{\downarrow}$ of two STM images derived from the individual spin channels.

In order to establish that a surface-sensitive technique, such as STM, allows one to access bulklike properties of impurities, we proceed in several steps: 
First, it is demonstrated that H passivation of the Si$(111)$ surface enables imaging of impurity-induced wave functions almost undisturbed by surface effects. 
Second, we show that the magnetic moment of Cr, Mn, and Fe impurities near the H-passivated Si$(111)$ surface is the same as in bulk, and that the energetic position of their electronic impurity states relative to the band edges is essentially unaltered compared to bulk. 
Third, by using Fe impurities as an example, it is found that the magnetic exchange interactions of subsurface TM impurities 
are very similar to those between impurities in bulk.

\begin{figure}[t]
	\centering
	\includegraphics[]{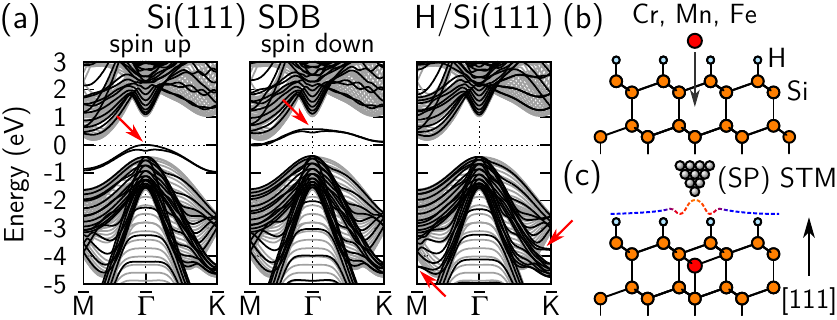}
	\caption{(a) Surface band structure of the Si$(111)$ single dangling bond termination showing surface states (red arrows) and their removal via H passivation. Gray lines correspond to the projected bulk band structure. (b) Deposition of (interstitial) TM impurities below the H layer and (c)~subsequent (SP)~STM analysis.}
	\label{fig:SiSurfaceStates}
\end{figure}

\begin{figure}[t]
	\centering
	\includegraphics[]{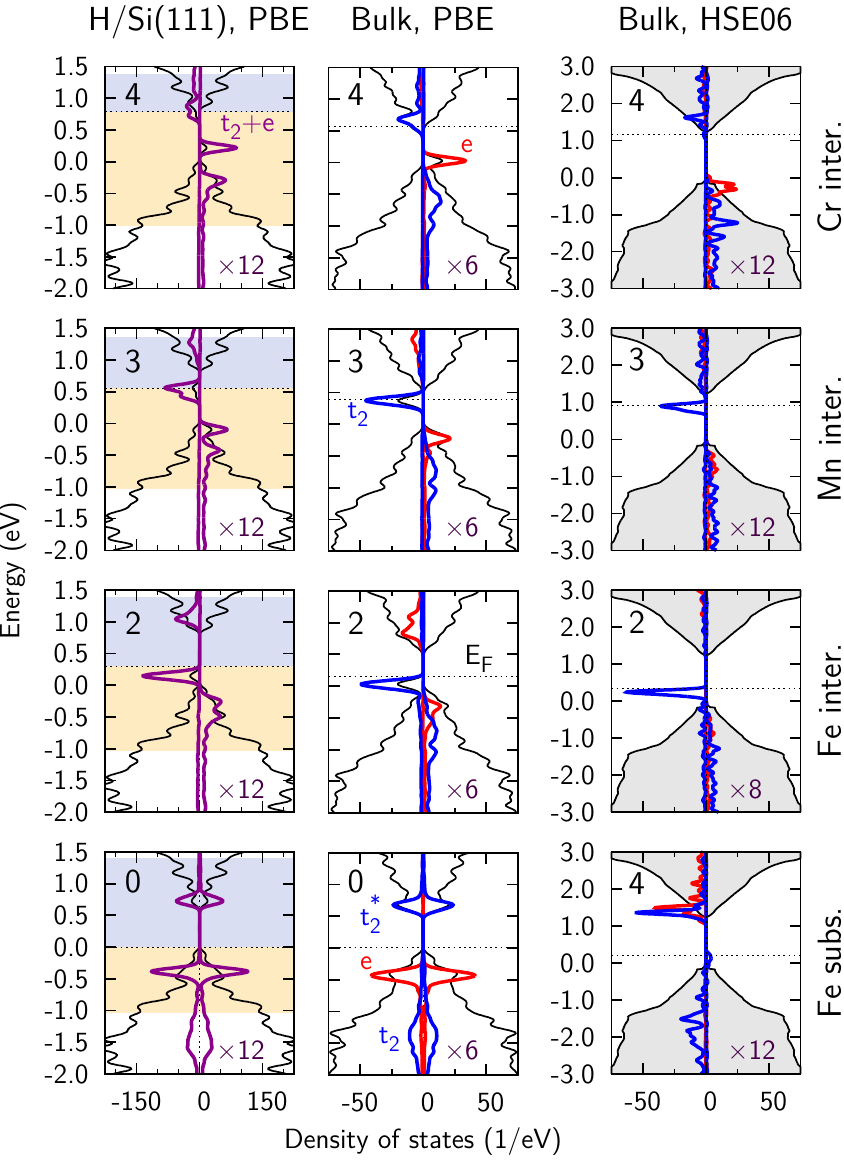}
	\caption{Spin-resolved electronic DOS for interstitial Cr, Mn, Fe, and substitutional Fe impurities. Compared are H/Si$(111)$ subsurface PBE [cf.~Fig.~\ref{fig:SiSurfaceStates}(c)], bulk PBE, and bulk HSE06 results. Thick, colored lines depict projections (scaled by the factors printed in the bottom-right corners of the frames) onto the TM $3d$ orbitals. The numbers in the top-left corners are the magnetic moments ($\mu_{\text{B}}$). Orange (blue) shaded areas indicate the filled-state (empty-state) energy integration intervals for the STM images. For each column a different projection technique and/or $\vec{k}$-point grid has been used. The gray shaded area in the HSE06 frames depicts the properly aligned total DOS of bulk Si.}
	\label{fig:DOS}
\end{figure}

Most commonly, STM is applied to semiconductor surfaces to map the wave functions of electronic surface states.
If these states are located in the fundamental band gap, they will dominate the images and thus preclude gathering of information about subsurface species.
Exceptions that are exploited in cross-sectional STM are some cleavage surfaces, e.g., GaAs$(110)$, where the STM image is dominated by states derived from the bulk valence band (VB) or conduction band (CB) edges.
In this work, we point out that, on most semiconductor surfaces, obstructing surface states can be removed by a suitable chemical passivation of the surface~\cite{Higashi-HSi:90}; e.g., on Si$(111)$, H passivation shifts the dangling-bond state of the surface Si atoms to much lower energies~\cite{Galli:10}, resulting in an insulating surface [cf.~Fig.~\ref{fig:SiSurfaceStates}(a)].
Such a passivation allows the STM to gather information about impurity-induced changes of the electronic structure.

The physics of \textit{isolated} $3d$ TM impurities in Si is rather well understood both experimentally and theoretically:
In agreement with established knowledge~\cite{Zhang:08,Shaughnessy:10},
we find for all three species, Cr, Mn, and Fe, the interstitial~(I) site to be energetically preferred over the substitutional~(S) site.
The $3d$ orbitals of Cr, Mn, and Fe give rise to electronic states in the Si band gap.
These levels are split into two groups of $t_2$ and $e$ symmetry due to the crystal field caused by the neighbor Si atoms, as well as shifted due to electronic exchange~\cite{Beeler:90}.
The ground-state magnetic moments resulting from our calculations can be rationalized by distributing the available valence electrons of each species among the exchange-split $t_2$ and $e$ levels.
In order to check the stability of the impurity magnetic moments and to avoid trapping in metastable states, we additionally performed constrained total magnetic moment calculations~\footnote{While Fe is numerically quite robust, Mn and Cr tend to converge to metastable magnetic states.}.
Our lowest-energy results agree with the established experimental and semilocal DFT values $4~\mu_{\text{B}}$, $3~\mu_{\text{B}}$, and $2~\mu_{\text{B}}$ (I) and $2~\mu_{\text{B}}$, $3~\mu_{\text{B}}$, and $0~\mu_{\text{B}}$ (S) for Cr, Mn, and Fe, respectively.
The other magnetic states are at least $324$~meV~(I) or $160$~meV~(S) higher.
Our calculations show that the magnetic moments remain unchanged in the proximity of a H/Si$(111)$ surface. 
Moreover, the \textit{whole} electronic structure is very similar for subsurface and bulk impurities (cf.~Fig.~\ref{fig:DOS}).
This very important finding provides the justification for the presented approach.
Test calculations without surface passivation led to strongly modified impurity magnetic moments.

\begin{figure*}[t]
	\centering
	\includegraphics[]{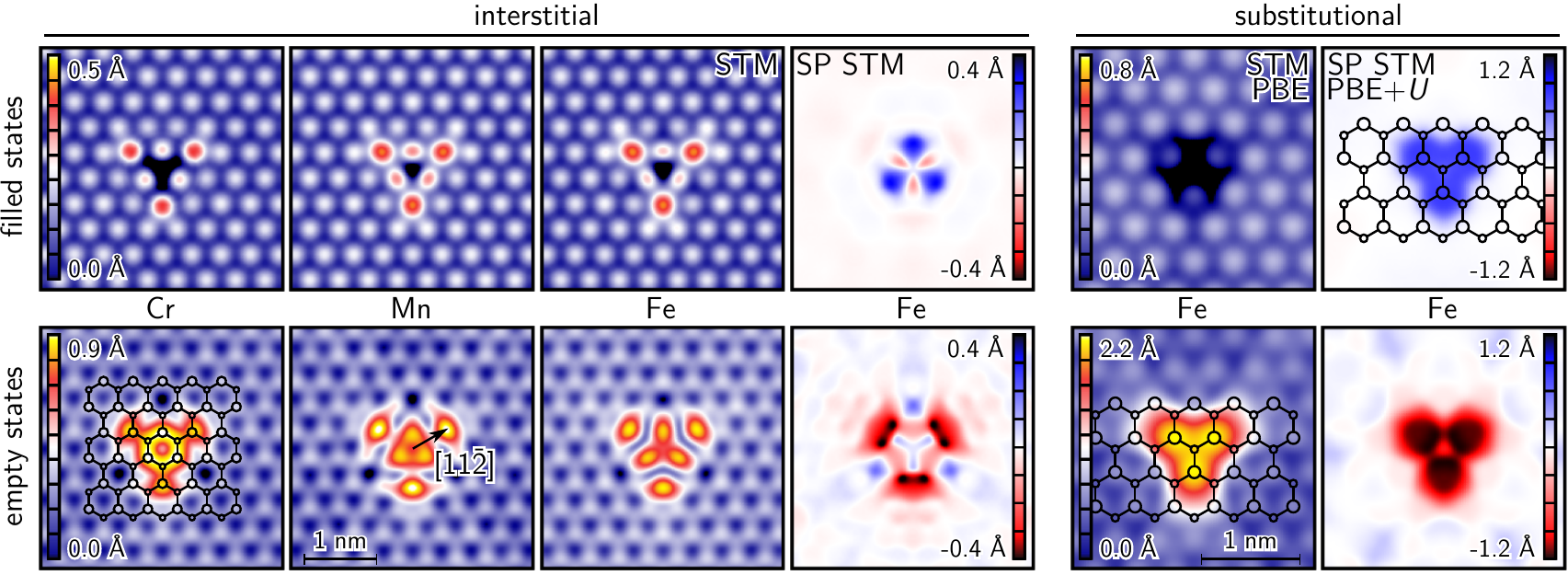}
	\caption{Simulated filled- and empty-state (SP) STM images for isolated interstitial and substitutional TM impurities (located in the center of each image) directly below the H/Si$(111)$ surface [cf.~Fig.~\ref{fig:SiSurfaceStates}(c)]. In the overlaid atomic structure, bigger circles mark the H/Si layers and smaller circles the second Si layer. For Fe$_\text{Si}$, PBE STM ($0~\mu_{\text{B}}$) and PBE$+U$ SP~STM ($4~\mu_{\text{B}}$) results are shown ($U_{3d}=3$~eV).}
	\label{fig:STM-isol}
\end{figure*}

While the $3d$-derived states of subsurface impurities are too localized to be imaged directly by STM,
the modification of host states near the band edges caused by the (spin-dependent) impurity potential
induces characteristic changes of the electronic structure.
These may be mapped by STM and used as a fingerprint for a specific impurity species.
One requirement to correctly predict these impurity-induced states
is the correct position of the $t_2$ and $e$ states \textit{relative} to the host band edges.
We therefore performed calculations with the hybrid functional HSE06 that yields a much more realistic size of the band gap in Si than the PBE functional~\footnote{Note that, since these calculations are excessively demanding, we used the optimized PBE atomic structure.}. 
For the interstitials (the three topmost rows in Fig.~\ref{fig:DOS}), the ground-state magnetic moments and the positioning of the impurity states in the (two times larger) band gap are very similar to the PBE case.
In particular, the position of the minority $t_2$ state (close to the VB for Fe, somewhat below the CB for Mn, touching the CB for Cr) is qualitatively similar in both functionals.
These similarities give confidence that the assignment of states to the
energy interval of filled-state versus empty-state images (cf.~Fig.~\ref{fig:DOS}, left column)
is independent of the functional.
Hence, already the frequently employed semilocal PBE functional~\cite{Zhang:08, KuewenBechstedt:09, Shaughnessy:10, Qian:06, Wu:07, Otrokov:11} which we will use for the simulations of STM images allows us to make verifiable predictions.
The last row of Fig.~\ref{fig:DOS} displays the DOS of substitutional Fe$_\text{Si}$.
Interestingly, there are clear differences in this case:
While PBE yields zero spin, the HSE06 functional (and also the PBE$+U$ method, $U_{3d}=3$~eV \cite{QE-LDA-U:05,KuewenBechstedt:09}, not shown) prefer a high-spin state.
Experimental information about the spin state of Fe$_\text{Si}$, as could be obtained by SP~STM (see below), would be very valuable.

Simulated STM images for isolated TM impurities are shown in Fig.~\ref{fig:STM-isol}.
The filled-state images of the interstitials show triangular features with little element specificity.
While their apparent height in the center, directly above the TM impurity, is reduced, the corners of the triangle are formed
by bright Si-H groups located in the $[11\bar{2}]$, $[\bar{2}11]$, and $[1\bar{2}1]$ directions from the center.
More interesting are the empty-state images.
Their corrugation is twice as large despite the smaller bias voltage used (energy integration interval, cf.~Fig.~\ref{fig:DOS}).
The features induced by the subsurface impurities are clearly different from those in the filled-state case.
The Cr feature has a ringlike shape with additional protrusions at the H atoms in the $[11\bar{2}]$, $[\bar{2}11]$, and $[1\bar{2}1]$ directions.
In contrast, the Fe feature consists of a central region resembling a caltrop, which is sharply separated from the protrusions at the mentioned H atoms.
Mn appears visually as a mixture of the Cr and the Fe feature. 
These differences are caused by the different amounts of 
hybridization between $t_2$-like minority spin states and the CB states of Si (cf.~Fig.~\ref{fig:DOS}).
We also simulated STM images for interstitials deeper below the surface (not shown).
The rapid loss of corrugation underlines that the wave functions decay quickly further away from the impurity.
From the subsurface STM image (and also from bulk cross sections, which are not shown here) 
we estimate a range of $\sim 20$~\AA\ (cf.~Fig.~\ref{fig:STM-isol}).
In summary, the TM-impurity-induced states in Si are very different in 
shape and size from those of subsurface Mn in GaAs, judging
from the experimental STM images of GaAs$(110)$~\cite{KoenraadFlatte:11, Garleff:08, Garleff:10},  
but also from the two-dimensional plots of calculated bulk wave functions~\cite{Jancu:08}.
The electronic states derived from the VB or the CB of the Si host crystal are both 
less extended than the states of VB character induced by Mn in GaAs.
Consequently, an even higher level of TM doping is necessary to achieve a comparable wave function overlap
and thus potentially ferromagnetic coupling between adjacent impurities.
However, the achievable concentration is limited by the formation of
TM-acceptor complexes~\cite{Sanati:07}, TM clusters,
or (mostly nonmagnetic) TM-Si compounds~\cite{Geisler:12, Geisler:13}.
We note that our method is capable of resolving impurity complexes
(especially TM-H complexes that might occur)
due to the deviating symmetry of their STM signature (see the Supplemental Material).

Next, we point out the added value of \textit{spin-polarized} STM images of TM impurities. 
If the samples are paramagnetic, an external magnetic field may be required to align the magnetic moments in order to obtain a magnetic contrast.
Two exemplary applications of the method are discussed: 
detection of the existence of a magnetic moment, e.g., in the case of Fe$_\text{Si}$, and 
measurement of the magnetic exchange interactions between impurity pairs.
For the interstitials, we find that the magnetic contrast increases with the magnetic moment of the impurity, i.e., from Fe to Cr. 
The spin polarization induced in the substrate electronic structure is anisotropic,
as can be seen in the case of interstitial Fe$_\text{I}$ in Fig.~\ref{fig:STM-isol}.
For the Fe$_\text{Si}$ impurity, which is expected to occur only rarely due to its high formation energy,
the SP~STM offers the unique opportunity to decide about the debated issue of a high-spin state. 
Our approach allows for detection of the impurity site and the magnetic moment \textit{at the same time}.
Hence, it becomes possible to safely identify and characterize Fe$_\text{Si}$ even in the presence of the more abundant Fe$_\text{I}$ impurities~\footnote{The unambiguous assignment of the lattice site requires a thorough characterization of the STM image of the undisturbed H/Si$(111)$ surface, which should preferentially be done at both positive and negative bias voltage.}.
The high-spin state predicted by both the HSE06 functional and PBE$+U$ calculations, if it exists, is clearly detectable due to its high corrugation of $\pm 1.2$~\AA\ in the SP~STM images (Fig.~\ref{fig:STM-isol}, right).  

\begin{table}[t]
\begin{center}
\caption{\label{tab:EnergyDifferences}Energy differences $\Delta E = E_{\text{AFM}} - E_{\text{FM}}$ ($\pm 0.5$ meV) per Fe atom and distances of two interacting Fe$_\text{I}$ impurities in the H/Si$(111)$ subsurface layer [for the geometry, cf.~Figs.~\ref{fig:SiSurfaceStates}(c) and~\ref{fig:STM-Fe-WW}] and in bulk Si. 
Configuration~D, whose $\vert \Delta E \vert$ value is already below the numerical accuracy, is included to indicate the finite interaction range.}
\begin{ruledtabular}
\begin{tabular}{lcccc}
Configuration & A & B & C & D	\\
\hline
Fe-Fe distance ($\AA$)			& $3.90$	& $6.70$	& $7.75$	& $10.22$		\\
$\Delta E$, H/Si$(111)$ (meV)	&  $+47$	&   $+3$	&   $-8$	& $-0.2$		\\
$\Delta E$, bulk Si (meV)		&  $+44$	&   $+2$	&   $-7$	& $-0.2$		\\
\end{tabular}
\end{ruledtabular}
\end{center}
\end{table}

\vspace{0.3cm}

\begin{figure}[t]
	\centering
	\includegraphics[]{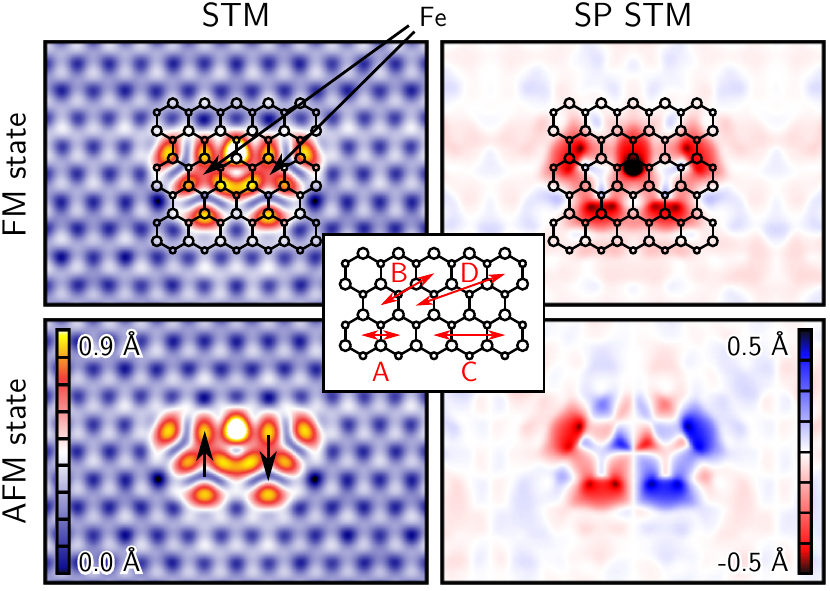}
	\caption{Simulated empty-state (SP)~STM images for two interacting Fe$_\text{I}$ impurities 
	in configuration C 
	directly below the H/Si$(111)$ surface [cf.~Fig.~\ref{fig:SiSurfaceStates}(c)] in two
	different magnetic states (FM/AFM).
	The overlaid atomic structure is the same as in Fig.~\ref{fig:STM-isol}.
	The inset illustrates other configurations, as reported in Table~\ref{tab:EnergyDifferences}.}
	\label{fig:STM-Fe-WW}
\end{figure}

Finally, we discuss how exchange coupling constants can be determined by the SP~STM method, given that
the impurity-impurity interactions are sufficiently weak so that the individual magnetic moment and local electronic structure
of each impurity are largely preserved.
The magnetic exchange interaction between TM impurities shows rich physics:
It may change both its magnitude and sign as a function of the distance vector of an impurity pair. 
We demonstrate this behavior by calculating the total energy difference $\Delta E$ between parallel [ferromagnetic (FM)] and antiparallel [antiferromagnetic (AFM)] local magnetic moment alignment for a pair of Fe interstitials (cf. Table~\ref{tab:EnergyDifferences}). 
FM interaction is found between neighboring Fe$_\text{I}$ (\enquote{A}). 
The exchange interaction is found to be strongly anisotropic:
For instance, the interaction along the $(11\bar{2})$ direction (\enquote{B}) is FM,
while the interaction between next-nearest-neighbor interstitials along the $(10\bar{1})$ direction is AFM,
despite the similar distance in both cases. 
Figure~\ref{fig:STM-Fe-WW} shows simulated STM images for configuration C. 
While conventional STM is predicted to see essentially a superposition of the two individual images, 
the calculated SP~STM contrast of $\pm 0.5$~\AA\ is sufficiently large to distinguish FM from AFM alignment. 
A comparison of bulk ($434$-atom cells) and subsurface $\Delta E$ for the example of two Fe$_\text{I}$
shows that not only the impurity-host but also the impurity-impurity interactions are \textit{quantitatively} preserved
in the vicinity of the passivated surface (cf.~Table~\ref{tab:EnergyDifferences}),
even though the TM impurities are maximally close to the surface here.
This remarkable agreement is, in the present case,
related to the rather short-ranged impurity-induced host states that mediate the interaction.
However, we point out that the capability to determine the size of exchange interactions
between bulk impurities by surface-sensitive SP~STM is not limited to this case: 
If the impurity-induced wave functions are spatially more extended,
this would allow the experimentalist to select an impurity pair further away from the surface that is still detectable.
Also in this case, the substrate contains a major fraction of the bulk impurity-induced wave function.
This ensures that the exchange interaction determined from SP~STM measurements still reflects the full size of the bulk exchange interaction.
One can now proceed with the following strategy:
After cleavage
of a grown sample with a random distribution of impurities
and subsequent passivation of the surface, one can select 
isolated pairs and measure the sign of their $\Delta E$ (FM/AFM) as a function of their distance vector.
Moreover, an external, gradually increased magnetic field can be used to measure $\Delta E$ by determining the Zeeman energy
at which the magnetic moments align parallel (switching), given that the ground state is AFM and that the interaction
is not too strong to be overcome by the magnetic field (e.g., \enquote{D} in Table~\ref{tab:EnergyDifferences}).

We stress that any semiconductor surface for which a passivation procedure is known is accessible to our proposed strategy. 
An intensely studied and controversially disputed system is Co-doped ZnO.
Up to now, the existence of FM order is an open question~\cite{Kittilstved:06,Ney:08}.
It has been shown recently that low-temperature exposition ($200$~K) of the ZnO$(10\bar{1}0)$ cleavage plane to H atoms
yields a fully passivated surface~\cite{Wang:05}.
Hence, by using the above strategy, the magnetic moment of
Co impurities and their distance-dependent interactions  
could be studied by STM and SP~STM, 
which would enable an assessment of their contribution to the magnetic properties of doped bulk ZnO. 

In summary, we demonstrated on the basis of \textit{ab initio} calculations how (SP)~STM can provide information about 
\textit{bulklike} impurity-host and impurity-impurity interactions  
below passivated semiconductor surfaces.
A comparison with hybrid functional results for Cr, Mn, and Fe interstitials in Si provided evidence that the semilocal PBE functional is 
sufficiently reliable for the simulation of STM images. 
We suggested an experimental route to resolve the issue of a high-spin versus low-spin ground state of Fe$_\text{Si}$,
which could become a benchmark for the applicability of hybrid functionals in the field of DMS.
Finally, we discussed how the SP~STM approach could be applied to DMS in general.

We thank M.~Gruyters and R.~Berndt, Christian-Albrechts-Universit\"{a}t zu Kiel, for discussions on experimental aspects.

\end{document}